\documentclass{llncs}
\usepackage{amssymb}
\usepackage{graphicx}
\usepackage{textcomp}

\begin{document}
\title{Ontology Based Data Integration Over Document and Column Family Oriented NOSQL stores}

\author{Olivier Cur\'e\inst{1}, Myriam Lamolle\inst{2},  Chan Le Duc\inst{2}}
\institute{Universit\'{e} Paris-Est, LIGM, Marne-la-Vall\'ee, France\\
\email{ocure@univ-mlv.fr}
\and
LIASD Universit\'e Paris 8 - IUT de Montreuil\\
\email{\{myriam.lamolle, chan.leduc\}@iut.univ-paris8.fr}
}

\maketitle
\begin{abstract}
The World Wide Web infrastructure together with its more than 2 billion users enables to store information at a rate that has never been achieved before.
This is mainly due to the will of storing almost all end-user interactions performed on some web applications.
In order to reply to scalability and availability constraints, many web companies involved in this process recently started to design their own data management systems.
Many of them are referred to as NOSQL databases, standing for 'Not only SQL'.
With their wide adoption emerges new needs and data integration is one of them.
In this paper, we consider that an ontology-based representation of the information stored in a set of NOSQL sources is highly needed.
The main motivation of this approach is the ability to reason on elements of the ontology and to retrieve information in an efficient and distributed manner.
Our contributions are the following: (1) we analyze a set of schemaless NOSQL databases to generate local ontologies, 
(2) we generate a global ontology based on the discovery of correspondences between the local ontologies and finally (3) we propose a query translation solution from SPARQL
to query languages of the sources.
We are currently implementing our data integration solution on two popular NOSQL databases: MongoDB as a document database and Cassandra as a column family store.

 

\end{abstract}

\section{Introduction}
The distributed architecture of the World Wide Web and its more than 2 billion users, in 2011, enables to store vast amount of information from end-user interactions. 
The volumes of data retrieved this way are so large that it motivated the design and implementation of new data models and management systems
able to tackle issues such as scalability, high availability and  partition tolerance.
In fact, the Web helped us to understand that the until now prevalent relational model does not fit all the data management issues \cite{DBLP:conf/icde/StonebrakerC05}.

These new data stores are regrouped under the NOSQL label (but coSQL \cite{DBLP:journals/cacm/MeijerB11} is another recently proposed name).
This acronym stands for 'Not Only SQL' and generally identifies data stores based on the Distributed Hash Table (DHT) model which provides a hash table access semantics.
That is in order to access and modify a data object, a client is required to provide the key for this object and a management system will lookup the object using an equality match to the required attribute key.
The First successful NOSQL databases were developed by Web companies like Google (with Big Table \cite{DBLP:conf/osdi/ChangDGHWBCFG06}) and Amazon (Dynamo \cite{DBLP:conf/sosp/DeCandiaHJKLPSVV07}).
An important number of open source projects followed more or less inspired by these two systems, e.g. MongoDB\footnote{http://www.mongodb.org/}, Cassandra\footnote{http://cassandra.apache.org/} which
respectively correspond to the document and column family categories.
Nowadays, NOSQL systems are used in all kinds of application domains (e.g. social networks, science, finance) and are present in cloud computing environments.
Hence, we consider that the Web of Data can not miss the opportunity to address and integrate technologies and datasets emerging from this ecosystem.

In this paper, we propose a data integration framework where the target schema is represented as a semantic web ontology and the sources correspond to NOSQL databases.
The main difficulty in integrating these data sources concerns their schemalessness and lack of a common declarative query language.

Concerning the schemalessness, although this provides for a form of flexibility in term of data modeling, this makes the generation of correspondences between a global and local schemata more involved.
Thus a first contribution of our work consists in generating a local schema for each integrated source using an inductive approach.
This approach uses non-standard description logic (DL \cite {baader:2003}) reasoning services like Most Specific Concept (MSC) and Least Concept Subsumer (LCS) in order to generate a concept for a group of similar individuals and 
to define hierarchies for these concepts.
Our second contribution enables the specification of a global ontology based on the local ontologies generated for each data source. 
This global ontology results from the correspondences discovered between concept definitions present in each local ontology.

Concerning the lack of a common declarative query language, we propose a Bridge Query Language (BQL) that supports a translation from SPARQL queries expressed over the
global ontology to the possibly different query languages accepted at the sources.
In general, document and column family databases do not provide for a declarative query language like SQL.
They rather propose a procedural approach based on the use of specific APIs, for instance for the Java language.
Hence our last contribution is to present the main steps involved in this transformation and to provide a sketch of the BQL language.

This paper is organized as follows. In Section 2, we present related works in ontology based data integration. Section 3 provides some background knowledge on NOSQL databases, non-standard DL reasoning services
 and some alignment methods.
Section 4 details our contributions in the design of our ontology-based data integration system and thus provides for an overview of this system's architecture.
In Section 5, we present the query processing solution adopted for our system.
Section 6 concludes the paper and gives perspectives on future works.

\section{Related work}
To the best of our knowledge, this paper is a first attempt to integrate data stored in NOSQL systems into an ontology based framework.
Hence, in this section, we focus on the broader subject of ontology-based data integration and concentrate on solutions addressing the relational model.
Most of the work dedicated to bridging the gap between ontologies and relational databases concentrated on 
defining mapping languages, query answering and its relationship with reasoning over the ontology.

MASTRO \cite{DBLP:journals/semweb/CalvaneseGLLPRRRS11} is  the reference implementation for the Ontology-Based Data Access (OBDA) approach.
In OBDA, ontologies, expressed in Description Logics, represent the conceptual layer of the data stored in relational databases.
It allows for both sound and complete conjunctive query answering over an ontology by retrieving data from a relational databases.
Most of the nice properties of MASTRO come from the computational characteristics of DL-Lite which motivated the creation of OWL2QL, an OWL2 fragment.
Nevertheless, MASTRO requires that both the global ontology and relational schemata are known in order to define semantic mappings.

In \cite{DBLP:journals/ws/DolbyFKSS09}, the SHER system is presented as a system for scalable conjunctive query answering over $\mathcal{SHIQ}$ ontologies where the ABox 
is stored in a relational database management system. A main contribution of this work is to implement an ABox \textit{summarization} technique which improves the computational performances of query answering.

In the Maponto tool \cite{DBLP:conf/otm/AnBM05}, the authors propose a solution that enables to define complex mappings from simple correspondences. 
This approach expects end-users or an external software to provide mappings and then uses them to generate new ones.
Maponto is being provided with a set of relational databases and an existing ontology.

Systems like MARSON \cite{DBLP:conf/semweb/HuQ07} and RONTO \cite{ronto} discover simple mappings by classifying the relations of a database schema and validate the mapping consistency that have
 been generated. Like Maponto, these systems require that the target ontology is provided.

In comparison with these systems, our approach deals with the absence of a schema at the sources and of global ontology.
Moreover, while all systems based on a relational model benefit from the availability of SQL, the existence of a common query language for the sources can not be assumed in the context 
of NOSQL databases.

\section{Background}
In this section, we present background knowledge concerning the two NOSQL databases we are focusing on in this paper, namely document and column-oriented stores.
This is motivated by their ability to provide an efficient solution to the scalability issue by enabling to scale out quickly.
For both of these approaches, we model a similar use case dealing with the submission and reviewing process of scientific conferences.
Concerning ontology related operations, we present non-standard reasoning services encountered in DL, 
i.e. MSC, LCS and GCS, and provide information on methods used to align expressive ontologies. 

\subsection{Document oriented databases}
Document oriented databases correspond to an extension of the well-known key-value concept where in this case the value consists of a structured document. 
A document contains hierarchically organized data similar to XML and JSON. 
This permits to represent one-to-one as well as one-to-many relationships in a single document. 
Therefore a complex document can be retrieved or stored without using joins. 
Since document oriented databases are aware of stored data, it enables to define document field indexes as well as to propose advanced query features. 
The most popular document oriented databases are MongoDB (10gen) and CouchDB (Apache).  

\begin{description}
\item[Example 1]
This document database (denoted \texttt{docDB}) stores data in 2 collections, namely \texttt{Person} and \texttt{Document}. 
In the \texttt{Person} collection, documents are identified by the email address of the person and contains information regarding the last name, first name, url, university, person type (i.e. either a user, author, conference member 
or reviewer) and possibly a list of reviewed document identifiers. 
The documents in the \texttt{Document} collection are identified by a 'doc' prefix followed by a unique numerical value. For each document, the system stores the title, email of the different authors (corresponding 
to keys in the \texttt{Person} collection), the abstract and full content of the paper. Finally, a list of reviews is stored for each document.
Fig. \ref{documentDB} presents a graphical representation of a document for each collection.
In this database, the reviews of a paper are stored within the paper document. This is easily structured in a document store which generally supports the nesting of documents.
Similarly, the documents a person needs to review are stored in \texttt{Person} documents, i.e. in \texttt{writeReview}.
\begin{figure}[h]
  \centering
  \includegraphics[scale=0.5]{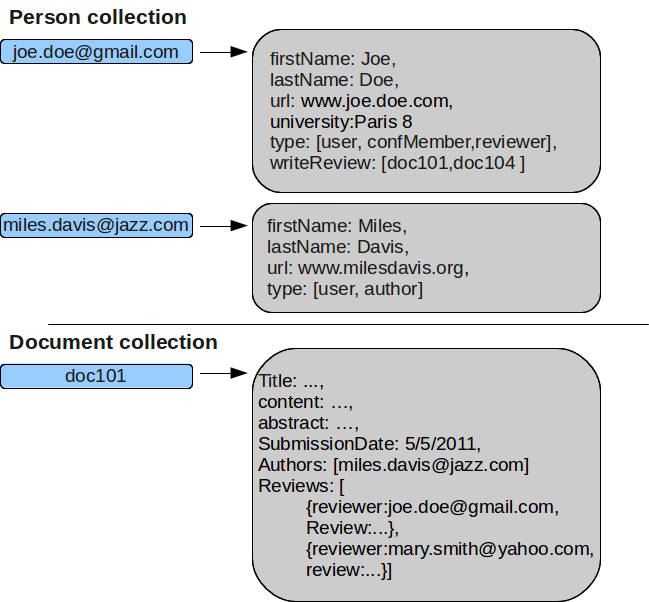}\\
  \caption{Extract of the document oriented database}
\label{documentDB}
\end{figure}
\end{description}

\subsection{Column-family databases}
Column family stores correspond to persistent, sparse, distributed multilevel hash maps. 
In column family stores, arbitrary keys (rows) are applied to arbitrary key value pairs (columns). 
These columns can be extended with further arbitrary key value pairs. 
Afterwards, these key value pair lists can be organized into column families and keyspaces. 
Finally, column-family stores can appear in a very similar shape to relational databases on the surface. 
The most popular systems are HBase and Cassandra. All of them are influenced by Google's Bigtable.

\begin{description}
\item[Example 2]
Considering the kind of queries one can ask on this column family (denoted \texttt{colDB}), the structure consists of 3 columns families: \texttt{Person}, \texttt{Paper} and \texttt{Review}. 
The set of information stored in these column families is the same as in Example 1.
The row key for the \texttt{Person}, \texttt{Paper} and \texttt{Review} are respectively the email address of the person and system generated identifiers for papers and reviews.
All other information entries are stored in columns with some of them being multi-valued. 
Fig. \ref{colDB} provides a graphical representation of an extract of \texttt{colDB}.
The \texttt{Paper} and \texttt{Review} column families have several columns in common (\texttt{abstract}, \texttt{content} and \texttt{submissionDate}).
But while the \texttt{authors} column stores the list of authors of a paper, the \texttt{author} column of reviewer stores the identifier of its reviewer.
\begin{figure}[h]
  \centering
  \includegraphics[scale=0.5]{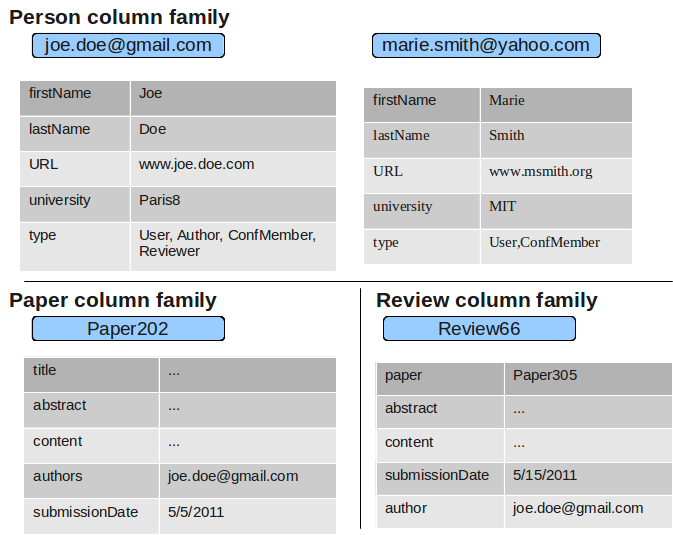}\\
  \caption{Extract of the column family database}
\label{colDB}
\end{figure}
\end{description}

\subsection{Non-standard Reasoning Services}
We present in this section the basic versions of \textit{Least Common Subsumer} (LCS) \cite{moller:1998} 
, \textit{Most Specific Concept} (MSC) \cite{baader:2003} and \textit{Good Common Subsumer} (GCS) \cite{baader:2004}.
The \textit{MSC} of an individual consists in defining the least concept description that the individual is an instance of.
\begin{description}
\item[Definition 1]
  Given concept terms $C_{1}, ..., C_{n}$, the \textit{MSC} of an individual $a$ is a concept term $C$ iff
\begin{itemize} 
 \item $C \sqsubseteq C_{i}$, for $1 \leqslant i \leqslant n$ ;
 \item $C$ is the most specific concept term with this property, i.e., if $D$ is a concept term such that $C_{i} \sqsubseteq D$ for $1 \leqslant i \leqslant n$, 
       then $C \sqsubseteq D$.
\end{itemize}
\end{description}
 
And, the \textit{LCS} of a set of concepts is the least concept that subsumes all of them, i.e., there is no sub-concept of this \textit{LCS} that
subsumes the set of concepts too.
\begin{description}
\item[Definition 2]
  Given concept terms $C_{1}, ..., C_{n}$, the LCS of $C_{1}, ..., C_{n}$ is a concept term $C$ such that
\begin{itemize}
 \item $C_{i} \sqsubseteq C$ for $1 \leqslant i \leqslant n$ ;
 \item $C$ is the least concept term with this property, i.e., if $D$ is a concept term such that $C_{i} \sqsubseteq D$ for $1 \leqslant i \leqslant n$, 
       then $C \sqsubseteq D$.
\end{itemize}
\end{description}

But, the \textit{LCS} is very hard to process in practice. 
So, Baader \cite{baader:2004} proposes an algorithm named \textit{Good Common Subsumer (GCS)} to compute an approximation of \textit{LCS} 
by determining the smallest conjunction of (negated) concept names subsuming the conjunction of the top level concept names of each considered concept.
By computing the \textit{MSC} and \textit{LCS} of these individuals, 
more complex concept descriptions can be added to the ontology \cite{baader:1999}.

\subsection{Alignment methods}
The heterogeneity between ontologies must be reduced in order to facilitate  
interoperability of applications based on these ontologies.  For this purpose, 
semantic correspondences between different entities belonging to two different ontologies are required 
to be established. This is the goal of ontology alignment as presented in \cite{euzenatShvaiko:2007}. 
An alignment consists of a set of correspondences between pairs of ontology entities.
Two entities of each pair are connected by a semantic relation (e.g. equivalence, subsumption, incompatibility, etc.). 
Moreover, a similarity measure can be associated to each correspondence to specify its trust. 
Then, a set of correspondences (i.e. alignment) can be used to merge ontologies, migrate data or translate queries from one to another ontology.

In the literature, there are several alignment methods that can be categorized according to techniques employed to produce alignments.  
The most early methods are based on the comparison of linguistic expressions \cite{euzenat:ISWC04}. 
Another aligner presented in \cite{david:IJSWIS07} has taken into account annotations of entities 
defined in ontologies. More recently, the methods introduced in \cite{djoufak:CAL08}, 
\cite{zghal:OM2008}, \cite{bach:2006} have exploited ontological structures related to concepts 
in question. These methods, namely simple alignment methods, are the most prevalent at present. 
They detect simple correspondences between atomic entities (or simple concepts) (e.g. $Human \sqsubseteq Person, Female \sqsubseteq Person$). 
As a result, some kinds of semantic heterogeneity in different ontologies can be solved by using these classical alignment methods.

However, simple correspondences are not sufficient to express relationships that represent correspondences between complex concepts since 
(i) it may be difficult to discover simple correspondences (or they do not exist) in certain cases, or 
(ii)  simple correspondences do not allow for expressing accurately relationships between entities.

A second important issue is that generating a complex alignment has a certain impact during the consistency checking of the system.
Indeed, a reasoner such as Pellet \cite{sir:JWS07} or FaCT++ \cite{TsHo06a}, running on a system consisting of two ontologies \textit{O$_{1}$} and \textit{O$_{2}$} 
and a simple alignment \textit{A$_{s}$}, may reply that the system is not consistent.
But, this same reasoner, with the same ontologies \textit{O$_{1}$} and \textit{O$_{2}$}, and with a complex alignment \textit{A$_{c}$} 
can deduce that the system is consistent.

Consequently, new works follow the way of complex alignment solutions such as \cite{ritze:ISWC09}. 
But, currently, they address the alignment of simple concept with a complex concept, at best.

\section{Architecture overview}
\begin{figure}[h]
  \centering
  \includegraphics[scale=0.5]{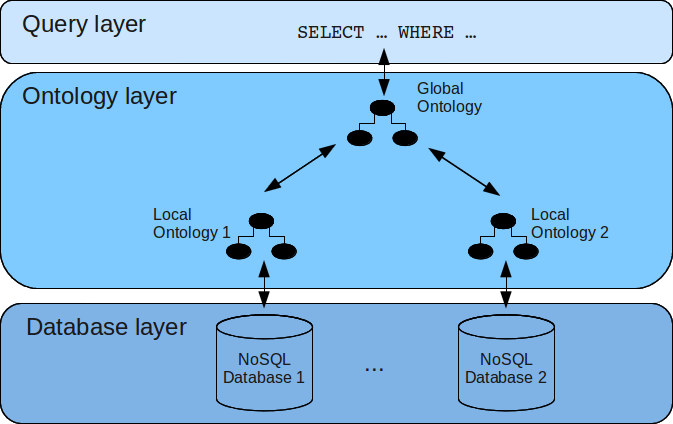}\\
  \caption{Basic architecture of our data integration system}
\label{diArchi}
\end{figure}

In this section, we present the main components of our system and highlight on the approaches used at each steps of the data integration processing.
These steps, depicted in Fig. \ref{diArchi}, consist of the (1) creation of an ontology associated to each data sources, (2) aligning these ontologies and (3) creating a global ontology given these correspondences.

Finally, we present a query language enabling to retrieve information stored in the sources from a query expressed over the global ontology.

\subsection{Source ontology generation}
As explained earlier, NOSQL databases are generally schemaless.
Although this provides flexibility for information storage, it makes the generation of associated ontologies more involved.
In fact, one can only use containers, i.e. collections and column families in respectively document and column family databases, of key/value pairs as well as key labels to deduce a schema.
Our approach considers that each container defines a DL concept and that each key label corresponds to a DL property that can either be a data type or object one
and whose domain is the DL concept corresponding to its container.

\begin{description}
\item[Example 3]
Consider Example 1 (resp. Example 2), the following concepts are automatically generated: \texttt{Person} and \texttt{Document} (resp. \texttt{Person}, \texttt{Paper} and \texttt{Review}).
Concerning DL properties, a \texttt{firstName} DL datatype property will be created in the cases of both Examples 1 and 2 with a domain corresponding to the \texttt{Person} DL concept.
Additionally, a \texttt{writeReview} DL object property is created with a domain and range corresponding to respectively the \texttt{Person} DL and \texttt{Paper} concepts.
This is due to the fact that the values of the \texttt{writeReview} (doc101 and doc 104 in the case of the document identified by joe.doe@gmail.com) served as identifier of other documents.
The same approach applies for \texttt{authors}, \texttt{author} and \texttt{paper} in Example 2.
\end{description}

A modeling pattern frequently encountered in key/value stores supports the discovery of complementary DL concepts and some subsumption relationships.
This pattern, henceforth denoted \textit{type definition}, consists of a key whose range of values is finite and which do not correspond to container identifiers, i.e. they do not serve as foreign keys.
We assume that each of these values specifies a DL concept.
For instance, this is the case of the \texttt{type} key in respectively Examples 1 and 2. Its set of possible values is \{User, Author, ConfMember and Reviewer\}, each of them corresponding to a DL concept.
These concepts can be organized into a hierarchy of concepts using methods of Formal Concept Analysis (FCA) \cite{DBLP:books/daglib/0095956}.
In a recent paper \cite{DBLP:journals/jcse/CureJ09}, we have emphasized on an FCA methodology for ontology mediation. 
Some features of this method are to create concepts that are not in the source ontologies, to label the new concepts, and to optimize the resulting ontology by eliminating redundant or irrelevant concepts. 
This approach easily applies to the discover of DL concepts and their subsumption relationships in the context of a \textit{type definition} pattern.
That is, tuples of the key of the pattern (\texttt{type} in our example) correspond to objects in the FCA terminology and their values provide FCA attributes. 
Then a Galois connection lattice can easily be computed using the methods proposed in \cite{DBLP:journals/jcse/CureJ09}.
The nodes of this lattice correspond to DL concepts and arrows between them specify subsumption relationships. 

\begin{description}
\item[Example 4]
We consider the document database of Fig.\ref{documentDB}.
The document identified by key 'joe.doe@gmail.com' has several \texttt{type} values (User, ConfMember, Author and Reviewer) while the 
document identified by 'miles.davis@jazz.com' is only characterized by the User value.
Using the information coming from different documents, one can discover the following DL concept subsumptions:
$Author \sqsubseteq User$\\
$Reviewer \sqsubseteq User$\\
$ConfMember \sqsubseteq User$\\
\end{description}

The method we have presented so far can be applied recursively to embedded structures where the nested container is reified into an object.

At this point in the local ontology generation process, we have created an ontology that is no more expressive than RDFS. 
We consider that using induction over the instances of the source database, we can enrich the ontology and leverage its expressiveness to a fragment of OWL2, namely OWL2EL.
This is performed using the approach proposed in \cite{DBLP:journals/corr/abs-1107-2822} to compute the GCS wrt to local ontology computed.
This method exploits the TBox of the ontology and precomputes the conjunction of concept names using FCA.
One issue in this precomputation is to handle a possibly very large set of FCA objects.

Ganter's \textit{attribute exploration} interactive algorithm \cite{DBLP:books/daglib/0095956} is an efficient approach for computing an appropriate representation of a concept lattice that at certain stages asks contextualized questions 
to a domain expert.
Instead of relying on this interactive process, we propose other solutions that may be used to select a subset of relevant objects.
In some cases, the set of objects may be of a reasonable size, (e.g. fitting into main memory) and a complete analysis is possible.
Nevertheless, in many situations, due to the size of individual data, a complete analysis is not realistic and some heuristics need to be proposed.
The first naive approach one can think of is to randomly access a set of individuals. Apart from the hazardous results this approach could provide, it is not just
doable in hash table context where the key of the container needs to be known a priori.

A simple heuristic consists in considering that the most frequently accessed individuals are the most representative of the ontology to generate.
In order to discover this set, one can take advantage of the data store architecture, generally distributed over several servers and supervised by several tools such as load balancers.
Using logs generated by these tools enables to identify a subset of the individuals that are the most frequently accessed in the application.

Finally an incremental schema generation approach can be implemented. That is each time a tuple is inserted or modified, the system checks if some labels are being introduced or deleted into the schema.
This approach imposes that each update operation goes through this process. 

At the end of this step, using an inductive approach, we have created a schema for each NOSQL source.
The goal of this schema is twofold: it enables the creation of DL ontology which can be serialized into an OWL2 fragment (namely OWL2EL) and supports the definition between ontology entities (i.e.
DL concepts and properties) with elements of the NOSQL source (i.e. documents, column families, columns and keys).
Hence, the arrows linking the database and ontology layers of Fig. \ref{diArchi} have been generated.
The task of the next section is to generate a global ontology via the discovery of alignments between local ontologies.

\subsection{Discovering Alignments between ontologies and global ontology building}
\begin{figure}[h]
 \centering
 \includegraphics[scale=0.5]{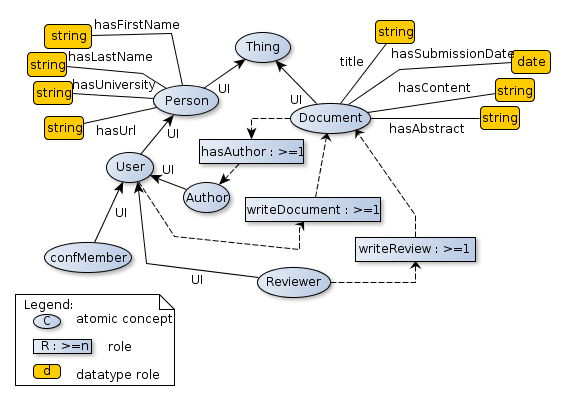}\\
 \caption{Extract of review document family DB graph}
\label{docDBgraph}
\end{figure}

We now propose a new solution to detect both simple and complex correspondences. 
To do this, we follow several steps. The first step consists in enriching the two ontologies to be aligned using the IDDL reasoner 
\cite{zimmermann:ICWRRS08}. This reasoner allows to add subsumption relations which are implicit in ontologies (see Fig.\ref{docDBgraph}).

The second step detects the simple correspondences using three classical alignment processes. 
We use the three conventional aligners OLA\footnote{OWL-Lite Alignment} \cite{djoufak:CAL08}, AROMA\footnote{Association Rule Ontology Matching
Approach}\cite{david:IJSWIS07} et WN\footnote{basic aligner of API alignment named JWNLAlignment}.
Each of them is based on a particular approach. The first aligner is a basic aligner, 
which uses the linguistic resource \textit{WordNet}, the second is based on a structural approach 
and the third on the annotations associated to entities. Note that the last two chosen aligners are considered by OAEI\footnote{Ontology Alignment Evaluation Initiative} 
among the best alignment systems.

The last step detects the complex correspondences. Our idea is inspired from simple alignment methods which are based on graphs 
\cite{djoufak:CAL08},\cite{zghal:OM2008}. Since, from a finite vocabulary, an infinite number of formulas
can be processed, it is impossible to know what are the relevant formulas to align.
A possible solution is to try to capture the semantic of OWL and to represent the constructors 
(for example, subsumption, disjunction, restriction of cardinality) by a graph formalism.
 Then, from the two graphs representing the ontologies to be aligned, it must search relevant subgraphs which can be aligned taking into account their structures 
and using a terminological similarity measure.

\subsubsection{Proposition 1}
Our first proposition allows correspondences between two complex concepts (i.e. formulas) to be detected. 

For example, $Paper~ \sqcap~ \exists write.Author~ \sqcap~ \exists accept.Reviewer~ \sqsubseteq~ Paper~ \sqcap~ \exists write.Author$ 
can be deduced.
The detection of this kind of matching is made by exploiting the structure of graphs, 
which expresses the semantics of the OWL-DL ontologies. 
Each graph consists of a set of subgraphs, 
which represent a formula. So, for all pairs of concepts ($C_{1}$, $C_{2}$) belonging to the ontologies ($O_{1}$, $O_{2}$), it is necessary to check
whether their respective subgraphs can be aligned.

A subgraph of a given concept \textit{C} consists of all concepts directly linked to \textit{C} 
by simple edges (subsumptions or disjunctions) or properties.

To align two subgraphs, one of the following cases must be checked:
\begin{enumerate}
 \item The first subgraph $SG_{1}$ subsumes the second subgraph $SG_{2}$ (i.e. $SG_{1} \sqsupseteq SG_{2}$). 
In this case, a relation of subsumption is generated;
 \item The second subgraph $SG_{2}$ subsumes the first subgraph $SG_{1}$ (i.e. $SG_{1} \sqsubseteq SG_{2}$). 
In this case, a relation of subsumption is generated, in the opposite way of case 1;
 \item The two subgraphs are equivalent. In other words, $SG_{1}$ subsumes $SG_{2}$ and $SG_{2}$ subsumes $SG_{1}$. 
In this case, a relation of equivalence is generated (i.e. $SG_{1} \equiv SG_{2}$).
\end{enumerate}

A subgraph $SG_{1}$ subsumes a subgraph $SG_{2}$ if the following conditions hold:
\begin{itemize}
 \item All direct subclasses of $SG_{1}$ are similar to direct subclasses of $SG_{2}$, 
 \item All disjoint subclasses of $SG_{1}$ are similar to disjoint subclasses of $SG_{2}$,
 \item All direct super classes of $SG_{1}$ or their generalization are similar to direct super classes of $SG_{2}$ 
or their generalization,
 \item All direct properties of $SG_{1}$ are similar to direct properties of $SG_{2}$, 
 \item All properties cardinalities of $SG_{1}$ are equivalent or subsumed by properties cardinalities of $SG_{2}$,
 \item All domains or co-domains of these properties of $SG_{1}$ are similar to domains or co-domains 
or their generalizations in $SG_{2}$.
\end{itemize}

\subsubsection{Proposition 2}
This proposition allow us to detect correspondences between a simple concept and a formula 
(e.g. $SubmittedPaper~ \sqsubseteq~ \exists submit.Author$). 
It is inspired by the research work presented in \cite{ritze:ISWC09} with some simplification and generalization. 
We search correspondences between simple concepts and formulas based on syntactic similarities 
between concepts and properties. It is necessary to use a purely syntactic similarity measure 
to compare concepts labels to properties labels. Moreover, a concept to align with a formula 
having a property similar syntactically must be a concept specializing a concept already aligned 
to a concept source or target of this property (or one of its super concept).

To generate the complex correspondences detected by these two propositions, 
we used the language EDOAL\footnote{http ://alignapi.gforge.inria.fr/edoal.html}, 
which extends the alignment format proposed by INRIA. This language can express complex structures 
between entities of different ontologies.

\subsubsection{Example}. 
\begin{sloppypar}
Given two ontologies $O_{1}$  and $O_{2}$ built from NOSQL databases to be aligned concerning a conference domain.
OWL semantics of these ontologies are represented by graphs  (cf. Fig. \ref{docDBgraph} of $O_{1}$ from example 1) .
 
The following simple correspondences are detected during the first step:
\begin{tabbing} \=  \\
\> $O_{1} : Document \equiv O_{2} : Document$	\\
\> $O_{1} : Person \equiv O_{2} : Person$	\\
\> $O_{1} : Reviewer \equiv O_{2} : Referee$	\\
\> $O_{1} : Review \equiv O_{2} : Review$	\\
\> $O_{1} : Conference \equiv O_{2} : Conference$	\\
\> $O_{1} : Submit \equiv O_{2} : Submit$	\\
\> $O_{1} : WriteReview \equiv O_{2} : WriteReview$	\\
\> $O_{1} : ConfMember \equiv O_{2} : ConfMember$
\end{tabbing}

\begin{figure}[t]
\begin{center}
\centering
 \includegraphics[scale=0.4]{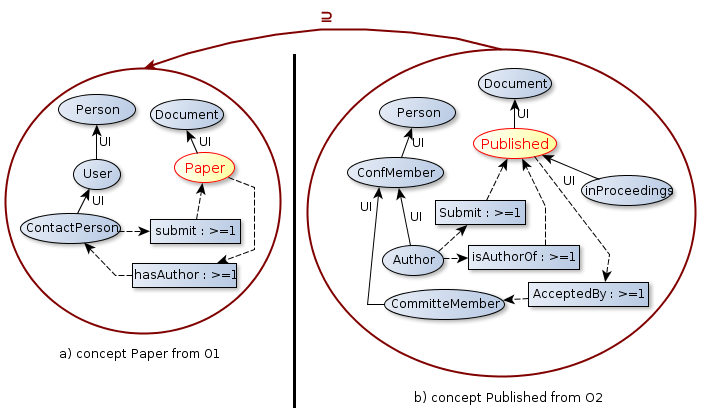}
 \caption{Aligned subgraphs of concepts $O_{1} : Paper$ and $O_{2} : Published$} \label{fig_ssgraph1}
\end{center}
\end{figure}

\begin{figure}[t]
\begin{center}
\centering
 \includegraphics[scale=0.4]{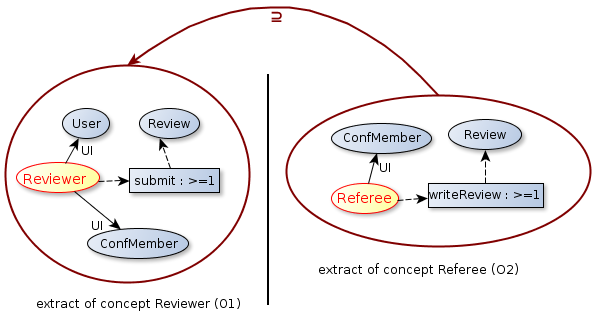}
 \caption{Aligned subgraphs of concepts $O_{1} : Reviewer$ and $O_{2} : Referee$} \label{fig_ssgraph2}
\end{center}
\end{figure}

The second step consisting in traversing the relevant subgraphs detects complex correspondences 
such as these presented in Fig. \ref{fig_ssgraph1} and \ref{fig_ssgraph2}. 
To do this, the neighborhood of the graphs nodes are considered. 
For example, the generated correspondence from the two subgraphs in Fig. \ref{fig_ssgraph1} 
having respectively the nodes $O_{1} : Paper$ and $O_{2} : Published$ as starting point is:

$O_{1} : Paper~ \sqcap~ \geqslant1~ O_{1} : hasAuthor.O_{1} : contactPerson~ \sqcap~ \geqslant1~ O_{1} : 
Submit \textasciimacron .O_{1} : contactPerson~ \sqsupseteq~ 
O_{2} : Published~ \sqcap~ \geqslant1~ O_{2} : Submit~ \geqslant ~ \textasciimacron.O_{2} : Author~ \sqcap~ 
\geqslant ~ O_{2} : isAuthorOf \textasciimacron .O_{2} : Author \sqcap~ O_{2} : AcceptedBy.O_{2} : ComitteMember$

In the same way, the generated correspondence from the two subgraphs of Fig. \ref{fig_ssgraph2} 
having respectively the nodes $O_{1}$ \textit{: Reviewer} and $O_{2}$ \textit{: Referee} as starting point is:

$O_{1} : Reviewer~ \sqcap~ \geqslant1~ O_{1} : WriteReview.O_{1} : Review~ \sqsubseteq~ 
O_{2} : Referee~ \sqcap~ \exists~ WriteReview.O_{2} : Review$ 

\end{sloppypar}

Taking into account the ontologies representing NOSQL data sources and their alignments, a global ontology \textit{GO} is built.
$GO=(O, A)$ represents networked ontologies $0= \left\lbrace 0_{1}, ..., O_{n} \right\rbrace$ 
through a set of alignments $A= \left\lbrace A_{1}, ..., A_{m} \right\rbrace$ where $A_{i}$ is the set of correspondences
between $0_{k}$ and  $0_{l} (k\neq l)$.

\section{Query processing}
In this section, we present the query processing solution adopted in our ontology-based data integration system.
The approach consists in two consecutive translation operations.

The first one transforms end-user written SPARQL queries expressed over the global ontology into a set of queries specified in the Bridge Query Language (BQL).
This translation uses the correspondences discovered during the local and global ontology generation steps and occurrences of a set of RDF/S properties (e.g. \texttt{rdf:type}, \texttt{rdfs:subClassOf}) in SPARQL queries. 
Given these correspondences, a BQL query is generated over the local ontology of a NoSQL source.
Due to space limitations, we do not provide a thorough presentation of BQL but rather sketch its main features.
BQL is a high-level declarative query language and has low-level, procedural programming flavor that enables to retrieve information from data repositories.
In fact, a BQL program is similar to specifying a query execution plan that can easily be translated into fully procedural programs satisfying a given API and programming language.
Following a nested data model, a BQL program specifies a sequence of steps that each define a single high level data operation.
Like a relational algebra, each step is specified via a relation definition which can serve as the input to another step.
A main construct of BQL is a \texttt{foreach .. in} operation which permits to iterate other a defined relation and perform some associated operations.
These operations generally consist in retrieving information from the database. 
This is specified using a \texttt{get} operation defined over a given database and container.
It contains 2 parameters: a set of filters expressed over source keys with standard comparator (e.g. =, $<$, $<=$, $\neq$) and a set of attributes to retrieve from the resultset.
In Example 5, we highlight a SPARQL to BQL transformation given our conference document database.

\begin{description}
 \item [Example 5]
Consider a query that retrieves titles of reviews written by a person with last name Doe.
This corresponds to the following SPARQL query which is simplified for readability reasons:\\
\texttt{SELECT ?t WHERE \{?p rdf:type Person. ?p hasLastName 'Doe'. \\
?p writeReview ?r. ?r hasTitle ?title.\}}\\
The presence of a \texttt{rdf:type} property in the SPARQL query provides some information about which source database and container we can create a BQL query for.
The query specifies that the ?p variable must of type \texttt{Person} which is mapped to the \texttt{person} container of the document NoSQL database.
This query addresses a list of reviews hence an iteration needs to be performed over the \texttt{writeReview} attribute of the \texttt{Person} container.
This first step of the BQL query is written as the following:\\
\texttt{temp(paper) = docDB.Person.get(\{lastName='Doe'\},\{writeReview\})}\\
Intuitively, the \texttt{temp} relation stores the list of review identifiers written by the person whose last name is 'Doe'.
The final result of the query is provided by the \texttt{ans} relation:\\
\texttt{ans(title) = foreach paper in temp : docDB.Paper.get(\{Key=paper\},\\
\{title\})}. That is for each identifier in the \texttt{temp} relation, find documents in the \texttt{Paper} collection of the docDB database and retrieve its title.
\end{description}

The second translation corresponds to generating a program in a given programming language (e.g. Java) from the different relations of a BQL query.
Given the procedural flavor of BQL, this translation is relatively straight forward but one set of rules needs to be defined for each language and each NoSQL database.
So far, we have implemented rules for the Java language for both the MongoDB and Cassandra stores.
In the future, we aim to define such rules for more NoSQL stores and programming languages (e.g. Python, Ruby).

\section{Conclusion}
This paper tackles the problem of integrating data stores in two of the most popular NOSQL database categories, i.e. document and column family oriented stores, in a Semantic Web 
context. It is well recognized that scalability is a main issue for these systems. 
The most involved aspect of this integration concerns the fact that these databases are schemaless and generally lack a common declarative query language.
Addressing this first issue, we emphasized that using existing techniques like FCA together with non-standard DL inferences like GCS, we could compute an ontology from the structure and instances 
of each databases source.
Using a novel alignment ontology method, we highlighted that these ontologies can be linked to create a global ontology over which SPARQL queries are expressed.
Finally, a bridge query language supports a translation approach to generate procedural queries, using specific APIs for each database source, from SPARQL queries.
We have already implemented this translation for the Java language for both the MongoDB and Cassandra NOSQL databases and we are currently working on query optimisation.
Recently, several propositions for a common NOSQL declarative query language are emerging (e.g. CQL for Cassandra, unQL for CouchDB). Studying these specifications is on our list of future works.
Nevertheless, we consider that our data integration framework is not complete until we incorporate another category of NOSQL stores: graph databases.

\bibliographystyle{abbrv}
\bibliography{obdiNosql}

\begin{thebibliography}{10}

\bibitem{baader:2003}
F.~Baader, D.~Calvanese, D.~L. McGuiness, D.~Nardi, and P.~Patel-Schneider.
\newblock {\em The Description Logic Handbook: Theory, Implementation,
  Applications}.
\newblock Cambridge University Press, Cambridge, UK, 2003.

\bibitem{baader:1999}
F.~Baader, R.~Küsters, and R.~Molitor.
\newblock Computing least common subsumers in description logics with
  existential restrictions.
\newblock pages 96--101. Morgan Kaufmann, 1999.

\bibitem{baader:2004}
F.~Baader, B.~Sertkaya, and A.~yasmin Turhan.
\newblock Computing the least common subsumer w.r.t. a background terminology.
\newblock In {\em Journal of Applied Logic}, pages 400--412. Springer, 2004.

\bibitem{DBLP:conf/osdi/ChangDGHWBCFG06}
F.~Chang, J.~Dean, S.~Ghemawat, W.~C. Hsieh, D.~A. Wallach, M.~Burrows,
  T.~Chandra, A.~Fikes, and R.~Gruber.
\newblock Bigtable: A distributed storage system for structured data (awarded
  best paper!).
\newblock In {\em OSDI}, pages 205--218, 2006.

\bibitem{david:IJSWIS07}
J.~David, F.~Guillet, and H.~Briand.
\newblock Association rule ontology matching approach.
\newblock {\em International Journal Semantic Web Information Systems},
  2:27--49, 2007.

\bibitem{DBLP:conf/sosp/DeCandiaHJKLPSVV07}
G.~DeCandia, D.~Hastorun, M.~Jampani, G.~Kakulapati, A.~Lakshman, A.~Pilchin,
  S.~Sivasubramanian, P.~Vosshall, and W.~Vogels.
\newblock Dynamo: amazon's highly available key-value store.
\newblock In {\em SOSP}, pages 205--220, 2007.

\bibitem{euzenat:ISWC04}
J.~Euzenat.
\newblock An api for ontology alignment.
\newblock {\em In 3rd conference on international semantic web conference
  (ISWC)}, pages 698--712, 2004.

\bibitem{euzenatShvaiko:2007}
J.~Euzenat and S.~Pavel.
\newblock {\em Ontology matching}.
\newblock Springer Verlag, Heidelberg (DE), 2007.

\bibitem{djoufak:CAL08}
J.-F. {Kengue Djoufak}, J.~Euzenat, and P.~Valtchev.
\newblock Alignement d'ontologies dirig\'e par la structure.
\newblock In Y.~A. Ameur, editor, {\em Conf\'erence Francophones sur les
  Architectures Logicielles, {CAL} 2008}, volume RNTI-L-2 of {\em RNTI}, pages
  155--. C\'epadu\`es-\'Editions, 2008.

\bibitem{bach:2006}
T.~{L\^e Bach}.
\newblock {\em Construction d'un Web s\'emantique multi-points de vue}.
\newblock Th\`ese de doctorat, \'Ecole des Mines de Paris \`a Sophia-Antipolis,
  2006.

\bibitem{moller:1998}
R.~Möller, V.~Haarslev, and B.~Neumann.
\newblock Semantics-based information retrieval.
\newblock In {\em In Int. Conf. on Information Technology and Knowledge
  Systems}, pages 48--61, 1998.

\bibitem{ritze:ISWC09}
D.~Ritze, C.~Meilicke, O.~\v{S}v\'{a}b Zamazal, and H.~Stuckenschmidt.
\newblock A pattern-based ontology matching approach for detecting complex
  correspondences.
\newblock {\em Proceedings of the ISWC 2009 Workshop on Ontology Matching},
  2009.

\bibitem{zghal:OM2008}
P.~Shvaiko, J.~Euzenat, F.~Giunchiglia, and B.~He, editors.
\newblock {\em SODA: an OWL-DL based ontology matching system}, volume 304 of
  {\em CEUR Workshop Proceedings}. CEUR-WS.org, 2008.

\bibitem{DBLP:conf/icde/StonebrakerC05}
M.~Stonebraker and U.~\c{C}etintemel.
\newblock "one size fits all": An idea whose time has come and gone (abstract).
\newblock In {\em ICDE}, pages 2--11, 2005.

\bibitem{zimmermann:ICWRRS08}
A.~Zimmermann and C.~{Le Duc}.
\newblock Reasoning with a network of aligned ontologies.
\newblock {\em In Proceedings of the 2nd International Conference on Web
  Reasoning and Rule Systems (ICWRRS)}, pages 43--57, 2008.

\end{thebibliography}

\end{document}